\documentclass{article}
\usepackage{amsmath,amssymb,bbm}

\newcommand{\asterisk}{*}
\newcommand{\tmop}[1]{\ensuremath{\operatorname{#1}}}
\newcommand{\tmtextbf}[1]{{\bfseries{#1}}}
\newcommand{\tmtextit}[1]{{\itshape{#1}}}
\newenvironment{itemizedot}{\begin{itemize} }{\end{itemize}}

\begin{document}

\title{A Fast Algorithm for the Discrete Core/Periphery Bipartitioning
Problem}\author{}\maketitle

\begin{center}
  Sean Z.W. Lip{\footnote{DAMTP, Centre for Mathematical Sciences, University
  of Cambridge, Wilberforce Road, Cambridge CB3 0WA, UK. E-mail address:
  S.Z.W.Lip@damtp.cam.ac.uk}}
\end{center}

\begin{abstract}
  Various methods have been proposed in the literature to determine an optimal
  partitioning of the actors in a network into core and periphery subsets.
  However, these methods either work only for relatively small input sizes, or
  do not guarantee an optimal answer. In this paper, we propose a new
  algorithm to solve this problem. This algorithm is efficient and exact,
  allowing the optimal partitioning for networks of several thousand actors to
  be computed in under a second. We also show that the optimal core can be
  characterized as a set containing the actors with the highest degrees in the
  original network.
\end{abstract}

\section{Introduction}

A concept that is prevalent in the field of social network analysis is the
core/periphery model. Such models arise in many fields of research, ranging
from corporate structure (Barsky, 1999) and world economics (Smith and White,
1992) to scientific citation networks (Mullins et al, 1977; Doreian, 1985) and
Japanese monkeys (Corradino, 1990).

As discussed in Borgatti and Everett (1999), a discrete core/periphery model
can be formulated as follows: consider a set of $n$ actors, labelled $1, 2,
\ldots, n$, and suppose that certain pairs of these actors interact. The idea
behind the model is that the actors can be partitioned into a cohesive
subgraph (a `core') and a loosely-connected `periphery'. A simple example is a
star graph, where the only ties that exist are those connecting a
distinguished node (1, say) to each of the other nodes. Then node 1 forms the
core, and the others form the periphery.

Several algorithms have been suggested for finding an optimal or near-optimal
decomposition of such a set into its core and peripheral parts. The simplest
approach is to try all possible subsets as the `core', and pick the one that
works best. However (as noted by Boyd et. al., 2006), there are exponentially
many such subsets, so this becomes infeasible quite rapidly as $n$ increases.
It therefore appears to be necessary to resort to heuristics, or prune the
search space in some way. Algorithms based on the former approach include the
genetic algorithm of Borgatti et. al. (2002) in the UCINET software package,
as well as algorithms based on simulated annealing and the Kernighan-Lin
algorithm, considered by Boyd et al (2006). An example of an algorithm which
prunes the search space can be found in the recent paper of Brusco (2011),
which develops an exact algorithm based on the branch-and-bound technique that
is feasible for networks with up to about 60 actors.

In this light, the main result of this paper might seem surprising: namely,
that it is possible to solve this problem exactly and efficiently, without
resorting to heuristics or pruning! This is true for both symmetric and
asymmetric networks. The solutions that will be described in this paper are
very fast, and therefore easily scalable to large networks. The basis of the
algorithm is a greedy procedure that systematically picks agents with maximal
degree to form part of the `core', and we will also prove that this algorithm
gives an optimal solution.

\section{Statement of the Problem}

We adopt a similar formulation to that used in Brusco (2011), and first
consider the case of symmetric networks (in which $A_{i j} = A_{j i}$ for all
$i$ and $j$). The symmetric core/periphery bipartitioning problem is defined
as follows:
\begin{itemizedot}
  \item There are $n$ actors, labelled $1, 2, \ldots, n$, and an $n \times n$
  binary adjacency matrix $A$ such that $A_{i j} = 1$ if actor $i$ interacts
  with actor $j$, and $A_{i j} = 0$ otherwise. (We do not consider
  self-interactions, and assume that, for each $i$, we have $A_{i i} = 0$.) .
  
  \item Define $S =\{1, \ldots, n\}$. We wish to find a proper, non-empty
  `core' subset $S_1 \subset S$ such that the following quantity is minimized:
  \begin{equation}
    \text{$Z (S_1) = \sum_{(i < j) \in S_1} \mathbbm{I}_{\{A_{i j} = 0\}} +
    \sum_{(i < j) \notin S_1} \mathbbm{I}_{\{A_{i j} = 1\}}$}
  \end{equation}
  (Here, we have employed the indicator function $\mathbbm{I}_{\{P\}}$, which
  is equal to 1 if the predicate $P$ is true, and 0 if $P$ is false.)
\end{itemizedot}
The intuitive idea behind this formulation is that we wish to maximize the
number of ties between actors in the core, and minimize the number of ties
between actors in the periphery. In an ideal scenario, there would be ties
between every pair of actors in the core, and no ties between any pair of
actors in the periphery. Notice that ties between core actors and periphery
actors do not appear in the expression for $Z (S_1)$; this is consistent with
the goal of Boyd et al. (2006) of finding a bipartition that simultaneously
maximizes connectivity in the core block and minimizes connectivity in the
periphery block.

\section{The Algorithm}

We now present a simple algorithm that solves the above problem in $O (n^2)$
time. Before doing this, however, we pause to make two definitions:
\begin{itemizedot}
  \item The \tmtextbf{degree} of a node $i$ is the number of ties incident to
  $i$. We represent this quantity by $\deg (i)$. It can be seen that $\deg (i)
  = \sum_{j \in S} a_{i j}$.
  
  \item Given a node $i$, and a subset $T \subseteq S$, we define $\delta_T
  (i)$ to be the number of ties joining $i$ with a node in $T$. In other
  words, $\delta_T (i) = \sum_{j \in T} a_{i j}$.
\end{itemizedot}
We now consider a restricted version of the problem, under the assumption that
the number of actors in the core, $S_1$, is fixed at the outset and is equal
to $k$ (where $1 \leqslant k < n$). There are therefore $\frac{k (k - 1)}{2}$
pairs of distinct actors in $S_1$, and each pair either has a tie between
them, or it does not. So we can write:
\begin{equation}
  \sum_{(i < j) \in S_1} \mathbbm{I}_{\{A_{i j} = 0\}} + \sum_{(i < j) \in
  S_1} \mathbbm{I}_{\{A_{i j} = 1\}} = \frac{k (k - 1)}{2} .
\end{equation}
Furthermore, the number of ties contributed by each node $i \notin S_1$ to the
periphery set is simply the degree of $i$, less the number of ties joining $i$
to a node in $S_1$. We can therefore write:
\begin{equation}
  \sum_{(i < j) \notin S_1} \mathbbm{I}_{\{A_{i j} = 1\}} = \frac{1}{2}
  \sum_{(i \neq j) \notin S_1} \mathbbm{I}_{\{A_{i j} = 1\}} = \frac{1}{2}
  \sum_{i \notin S_1} \left( \deg (i) - \delta_{S_1} (i) \right) .
\end{equation}
Using these two results, we can express $Z (S_1)$ as follows:
\begin{eqnarray}
  Z (S_1) & = & \sum_{(i < j) \in S_1} \mathbbm{I}_{\{A_{i j} = 0\}} +
  \sum_{(i < j) \notin S_1} \mathbbm{I}_{\{A_{i j} = 1\}} \nonumber\\
  &  &  \nonumber\\
  & = & \frac{k (k - 1)}{2} - \sum_{(i < j) \in S_1} \mathbbm{I}_{\{A_{i j} =
  1\}} + \frac{1}{2} \sum_{i \notin S_1} \left( \deg (i) - \delta_{S_1} (i)
  \right) \nonumber\\
  &  &  \nonumber\\
  & = & \frac{k (k - 1)}{2} - \frac{1}{2} \sum_{i \in S_1} \delta_{S_1} (i) +
  \frac{1}{2} \sum_{i \notin S_1} \deg (i) - \frac{1}{2} \sum_{i \notin S_1}
  \delta_{S_1} (i) \nonumber\\
  &  &  \nonumber\\
  & = & \left( \frac{1}{2} \sum_{i \in S} \deg (i) + \frac{k (k - 1)}{2}
  \right) - \frac{1}{2} \left( \sum_{i \in S_1} \deg (i) + \sum_{i \in S}
  \delta_{S_1} (i) \right)  \nonumber\\
  &  &  \nonumber\\
  & = & \left( \frac{1}{2} \sum_{i \in S} \deg (i) + \frac{k (k - 1)}{2}
  \right) - \sum_{i \in S_1} \deg (i) 
\end{eqnarray}
where the final equality arises because
\begin{equation}
  \sum_{i \in S} \delta_{S_1} (i) = \sum_{i \in S} \sum_{j \in S_1}
  \mathbbm{I}_{\{A_{i j} = 1\}} = \sum_{j \in S_1} \sum_{i \in S}
  \mathbbm{I}_{\{A_{j i} = 1\}} = \sum_{j \in S_1} \deg (j) .
\end{equation}
If $k$ is fixed, the terms in the first bracket are independent of the choice
of $S_1$, so our problem reduces to finding an $S_1$ of size $k$ such that the
final term is maximized. Clearly, we should therefore take $S_1$ to consist of
the $k$ nodes with largest degree in $S$. This can be done in $O (n \log n)$
time, since it takes $O (n \log n)$ time to sort the nodes in descending order
of degree using a standard algorithm such as merge sort (Knuth, 1998), and a
further $O (k)$ time to construct $S_1$.

We can now return to the original problem and treat the case in which $k$ is
unknown. Assume that the nodes are sorted in descending order of degree, and
that the resulting list of nodes is $\{v_1, v_2, \ldots, v_n \}$. We can then
determine the optimal $S_1$ by iterating through the possible values of $k$
and calculating the optimal $Z (S_1)$ for each, and finally taking the best
one.

Note that we need not repeat the calculation from scratch in each iteration,
because of the following observation: the addition of $v_k$ to the optimal set
increases the value of $Z$ by
\begin{equation}
  \left( \frac{k (k - 1)}{2} - \sum^k_{i = 1} \deg (v_i) \right) - \left(
  \frac{(k - 1) (k - 2)}{2} - \sum^{k - 1}_{i = 1} \deg (v_i) \right) = k - 1
  - \deg (v_k) .
\end{equation}
Initially, the core set is empty, and so the starting value of $Z$ is
\begin{equation}
  \text{$Z (\emptyset) = \sum_{(i < j) \in S} \mathbbm{I}_{\{A_{i j} = 1\}} =
  \frac{1}{2} \sum_{i \in S} \deg (i)$.}
\end{equation}
The full algorithm can therefore be specified as follows:
\begin{enumerate}
 \item Calculate and store the degrees of each node. Then sort the nodes in descending order of degree, to get a list of nodes $\{v_1, v_2, \ldots, v_n
\}$.
 \item Set $Z_{\tmop{best}}$ := $\infty$ and $k_{\tmop{best}}$ := $0$. (Note: instead of $\infty$, a suitably large upper bound, such as $n^2$, can be
used.)
 \item Set $Z$ := $\frac{1}{2} \sum_i \deg (i)$.
 \item For each $k$ from $1$ to $n - 1$, inclusive: set $Z$ := $Z + k - 1 - \deg (v_k)$. Then, if $Z < Z_{\tmop{best}}$, set $Z_{\tmop{best}}$ := $Z$ and
$k_{\tmop{best}}$ := $k$.
 \item Set $S_1$ := \{$v_1, \ldots, v_{k_{\tmop{best}}}$\}.
 \item Return $S_1$.
\end{enumerate}

Reading the input (i.e., the adjacency matrix describing the network) takes $O
(n^2)$ time, and so does calculating the degrees of each node. Sorting the
nodes takes $O (n \log n)$ time, and all other operations take $O (n)$ time,
so the algorithm runs in $O (n^2)$ time. If the input data is presented in the
form of an adjacency list (i.e., as a set of $n$ lists such that the $i$-th
list contains the neighbours of the $i ^{\tmop{th}}$ actor), or simply as a
list of existing ties, the algorithm would run in $O (n \log n + m)$ time,
where $m$ is the number of ties in the network.

This algorithm is therefore a significant improvement on both the
branch-and-bound and the heuristic approaches. The branch-and-bound method
provides an optimal answer, but is slow; the heuristic approaches do not
guarantee an optimal answer. The algorithm just described provides an optimal
answer, and does so quickly.

As an aside, it is possible to improve the main part of this algorithm
further. Let $Z_k$ be the value of $Z$ after the $k$-th iteration of the
algorithm. Notice that the sequence $\{k - 1 - \deg (v_k) : 1 \leqslant k <
n\}$ is non-decreasing, since the sequence $\{\deg (v_k) : 1 \leqslant k <
n\}$ is non-increasing. Therefore, there exists a $k^{\asterisk}$ such that
$Z_1 \geqslant Z_2 \geqslant \ldots \geqslant Z_{k^{\asterisk}} \leqslant
Z_{k^{\asterisk} + 1} \leqslant \ldots \leqslant Z_{n - 1}$. This observation
allows us to determine the optimum value of $k$ in $O (\log n)$ time, by
binary searching on $k$ to find the largest $k^{\asterisk}$ such that
$k^{\asterisk} - 1 - \deg (v_{k^{\asterisk}}) \leqslant 0$. Once we have found
this optimal value, we pick our core to be $S_1 =\{v_1, \ldots,
v_{k^{\asterisk}} \}$, as before. However, this does not lead to an
order-of-magnitude improvement in the time complexity, because, e.g., it still
takes $O (n \log n)$ time to sort the nodes at the beginning of the algorithm.

\section{Generalization to Asymmetric Networks}

In the version of the problem described by Brusco (2011), the underlying
networks were allowed to be symmetric or asymmetric. We now consider the
asymmetric case. The definition of the matrix $A$ then changes slightly: we
now have $A_{i j} = 1$ if there is a tie from actor $i$ to actor $j$, and
$A_{i j} = 0$ otherwise. The objective is now to find a proper subset $S_1
\subset S$ such that
\begin{equation}
  \text{$Z (S_1) = \frac{1}{2} \sum_{(i < j) \in S_1} \left(
  \mathbbm{I}_{\{A_{i j} = 0\}} +\mathbbm{I}_{\{A_{j i} = 0\}} \right) +
  \frac{1}{2} \sum_{(i < j) \notin S_1} \left( \mathbbm{I}_{\{A_{i j} = 1\}}
  +\mathbbm{I}_{\{A_{j i} = 1\}} \right)$}
\end{equation}
is minimized.

To solve this version of the problem, we introduce a symmetric weight function
$w (i, j) = \frac{1}{2} (\mathbbm{I}_{\{A_{i j} = 1\}} +\mathbbm{I}_{\{A_{j i}
= 1\}})$, for any two nodes $i \neq j$. Then we can write
\begin{equation}
  \text{$Z (S_1) = \sum_{(i < j) \in S_1} (1 - w (i, j)) + \sum_{(i < j)
  \notin S_1} w (i, j)$} .
\end{equation}
Finally, we redefine $\deg (i) = \sum_{j \in S} w (i, j)$, and $\delta_T (i) =
\sum_{j \in T} w (i, j)$. It is now straightforward to check that the analysis
in Section 3 carries over to this case, after we replace $\mathbbm{I}_{\{A_{i
j} = 0\}}$ with $(1 - w (i, j))$, and $\mathbbm{I}_{\{A_{i j} = 1\}}$ with $w
(i, j)$. Therefore, the algorithm in Section 3 still holds (albeit with a
modified definition of degree), and its time complexity remains unchanged.

\section{Tests of the Algorithm}

For input graphs with $n$ up to about 1000, the algorithm runs in under a
second. This could be sped up significantly if the graph is sparse ($m \ll
n^2$) and the data is presented in the form of an adjacency list (or a list of
ties), since the algorithm then takes $O (n \log n)$ time and can therefore
handle networks with $n$ up to about 50000 in under a second. (These estimates
are conservative.)

As a check, the algorithm described in Section 3 was tested, together with the
brute force algorithm (which tries every possible subset of $S$ as the core
and is therefore guaranteed to produce the optimal answer), on 100 random
input cases with $5 \leqslant n \leqslant 25$. Both algorithms produced the
same answer each time, and our algorithm is noticeably faster.

\section{Conclusions}

We have presented an exact, efficient algorithm to solve a discrete
core/periphery bipartitioning problem. This algorithm outperforms both the
heuristic and exhaustive search methods that have so far been used, and vastly
increases the sizes of the problems that can be tackled.

We also offer the qualitative insight that the actors which make up the core
are simply the ones with the most connections in the original network. As the
actors with highest degree are inserted into the core, the size of the core
increases until it hits a well-defined threshold. Beyond this threshold, it
becomes less attractive to add new actors to the core because the degrees of
the entering actors are not large enough to compensate for the core's
increasing size.

Note that this particular formulation of the core/periphery bipartitioning
problem is solved by choosing the most central nodes to lie in the core, where
`centrality' in this case is defined as degree centrality. However, other
measures of centrality are often used (Wasserman and Faust, 1994), and it may
be possible to formulate alternative definitions of a core/periphery
bipartitioning in which the optimal solution takes into account the
betweenness or closeness centralities of the actors. Furthermore, it would be
interesting to try and extend the algorithm presented in this paper to other
variants of the core/periphery bipartitioning problem, some of which have
continuous (as opposed to discrete) formulations.

\section{Acknowledgments}

The author is very grateful to the Gates Cambridge Trust for support.

\section{References}

\begin{itemizedot}
  \item Barsky, Noah P., 1999. A core/periphery structure in a corporate budgeting
process. Connections 22(2), 22--29.
  \item Borgatti, S.P., Everett, M.G., 1999. Models of core/periphery structures.
Social Networks 21, 375--395.
  \item Borgatti, S.P., Everett, M.G., Freeman, L.C., 2002. Ucinet for Windows:
Software for Social Network Analysis. Analytic Technologies, Harvard, MA.
  \item Boyd, J.P., Fitzgerald, W.J., Beck, R.J., 2006. Computing core/periphery
structures and permutation tests for social relations data. Social Networks 28
(2), 165--178.
  \item Brusco, M.J., 2011. An exact algorithm for a core/periphery bipartitioning
problem. Social Networks 33 (1), 12--19.
  \item Corradino, C., 1990. Proximity structure in a captive colony of Japanese
monkeys (\tmtextit{Macaca fuscata fuscata)}: an application of
multidimensional scaling. Primates 31 (3), 351--362.
  \item Doreian, P., 1985. Structural equivalence in a psychology journal network.
American Society for Information Science 36 (6), 411--417.
  \item Knuth, D., 1988. `Section 5.2.4: Sorting by Merging', The Art of Computer
Programming Volume 3. Addison-Wesley, 158--168.
  \item Mullins, N.C., Hargens, L.L., Hecht, P.K., Kick, E.L., 1977. The group
structure of cocitation clusters: a comparative study. American Sociological
Review 42, 552--562.
  \item Smith, D., White, D., 1992. Structure and dynamics of the global economy:
network analysis of international trade 1965--1980. Social Forces 70,
857--893.
  \item Wasserman, S., Faust, K., 1994. Social Network Analysis: Methods and
Applications. Cambridge University Press, 177--198.
\end{itemizedot}

\end{document}